\documentclass[traditabstract]{aa} 
\usepackage{txfonts}
\usepackage{graphicx}
\usepackage{longtable}

\usepackage{fixltx2e}
\begin{document}

\title{Two candidate brown dwarf companions around core helium-burning stars}

\author{V. Schaffenroth \inst{1,2} 
   \and L. Classen \inst{1}
   \and K. Nagel \inst{1}
   \and S. Geier \inst{3}
   \and C. Koen \inst{4}
   \and U. Heber \inst{1}
   \and H. Edelmann \inst{1}}


\institute{Dr. Karl Remeis-Observatory \& ECAP, Astronomical Institute,
Friedrich-Alexander University Erlangen-Nuremberg, Sternwartstr. 7, D 96049 Bamberg, Germany\\
\email{veronika.schaffenroth@sternwarte.uni-erlangen.de}
\and Institute for Astro- and Particle Physics, University of Innsbruck, Technikerstr. 25/8, 6020 Innsbruck, Austria
\and European Southern Observatory, Karl-Schwarzschild-Str.~2, 85748 Garching, Germany
\and Department of Statistics, University of the Western Cape, Private Bag X17, Bellville 7535, South Africa
}

\date{Received 16 July 2014\ Accepted 15 September 2014}

\abstract{Hot subdwarf stars of spectral type B (sdBs) are evolved, core helium-burning objects. The formation of those objects is puzzling, because the progenitor star has to lose almost its entire hydrogen envelope in the red-giant phase. Binary interactions have been invoked, but single sdBs exist as well. We report the discovery of two close hot subdwarf binaries with small radial velocity amplitudes. Follow-up photometry revealed reflection effects originating from cool irradiated companions, but no eclipses. The lower mass limits for the companions  of CPD-64$^{\circ}$481 ($0.048\,M_{\rm \odot}$) and PHL\,457 ($0.027\,M_{\rm \odot}$) are significantly below the stellar mass limit. Hence they could be brown dwarfs unless the inclination is unfavourable. Two very similar systems have already been reported. The probability that none of them is a brown dwarf is very small, 0.02\%. Hence we provide further evidence that substellar companions with masses that low are able to eject a common envelope and form an sdB star. Furthermore, we find that the properties of the observed sample of hot subdwarfs in reflection effect binaries is consistent with a scenario where single sdBs can still be formed via common envelope events, but their low-mass substellar companions do not survive.\\

}

\maketitle

\section{Introduction \label{s:intro}}

Hot subdwarf B stars (sdBs) are evolved, core helium-burning objects with only thin hydrogen envelopes and masses around $0.5\,M_{\rm \odot}$ 
(Heber \cite{heber86}, see Heber \cite{heber09} for a review). To form such objects, the progenitor star has to lose almost its entire hydrogen envelope in the red-giant phase. 

About half of the sdB stars are in close binaries with short periods from just a few hours to a few days (Maxted et al. \cite{maxted01}; Napiwotzki et al. \cite{napiwotzki04a}). Because the separation in these systems is much smaller than the size of the red-giant progenitor star, these binaries must have experienced a common-envelope and spiral-in phase (Han et al. \cite{han02,han03}). Although the common-envelope ejection channel is not properly understood (see Ivanova et al. \cite{ivanova13} for a review), it provides a reasonable explanation for the strong mass loss required to form sdB stars. However, for the other half of the known single-lined hot subdwarfs there is no evidence for close stellar companions as no radial velocity variations are found (Classen et al. \cite{classen11}).  

Soker (\cite{soker98}) proposed that substellar objects like brown dwarfs (BDs) and planets, which enter the envelope of a red giant, might be able to trigger its ejection. Substellar objects with masses higher than $\simeq10\,M_{\rm J}$ were predicted to survive the common envelope phase and end up in a close orbit around the stellar remnant, while planets with lower masses would evaporate or merge with the stellar core. The stellar remnant is predicted to lose most of its envelope and settle on the extreme horizontal branch (EHB). Such a scenario has also been proposed to explain the formation of single low-mass white dwarfs (Nelemans \& Tauris \cite{nelemans98}). 

\begin{figure*}[t!]
\centering
	\resizebox{9.15cm}{!}{\includegraphics{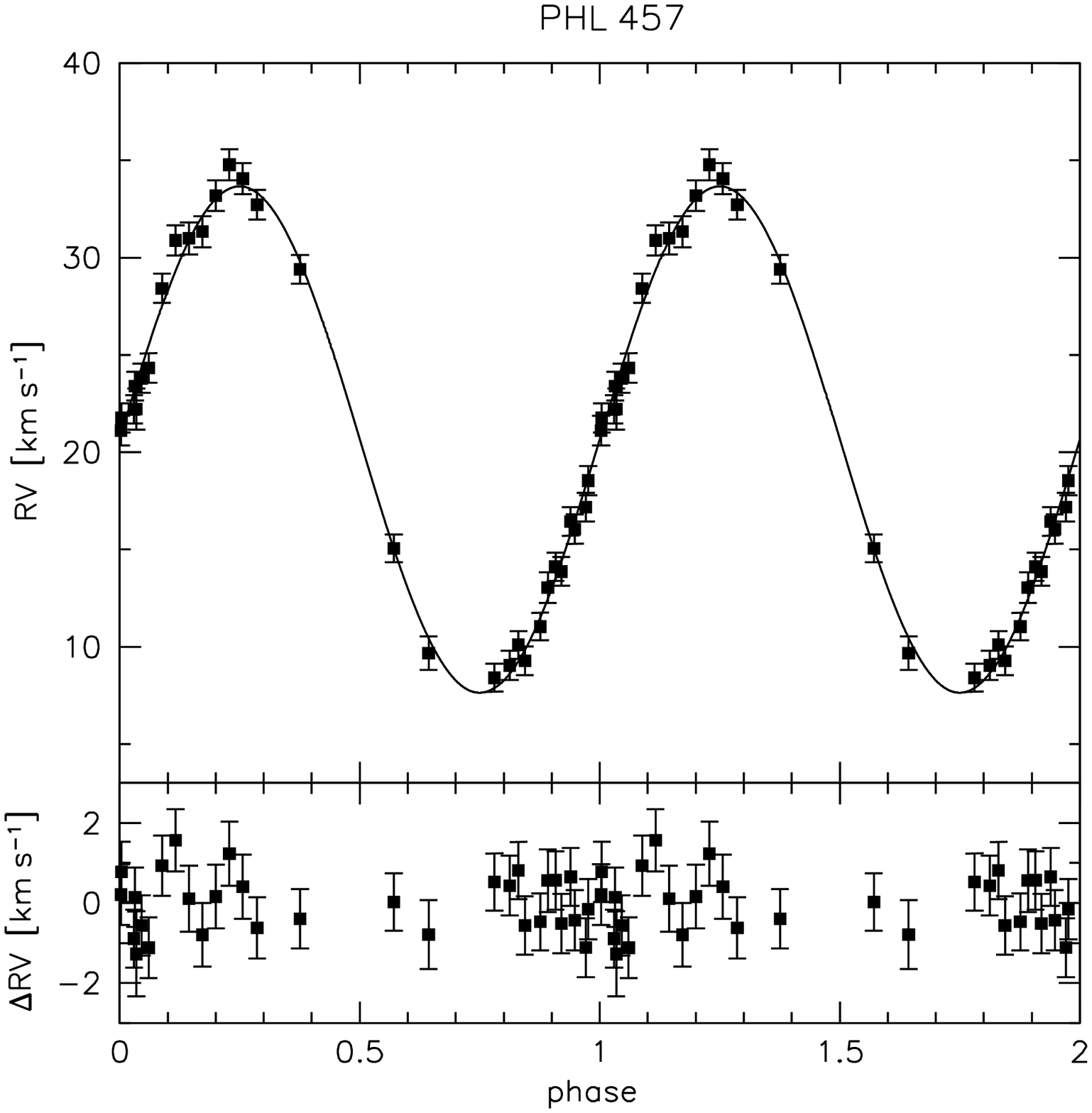}}
        \resizebox{9.15cm}{!}{\includegraphics{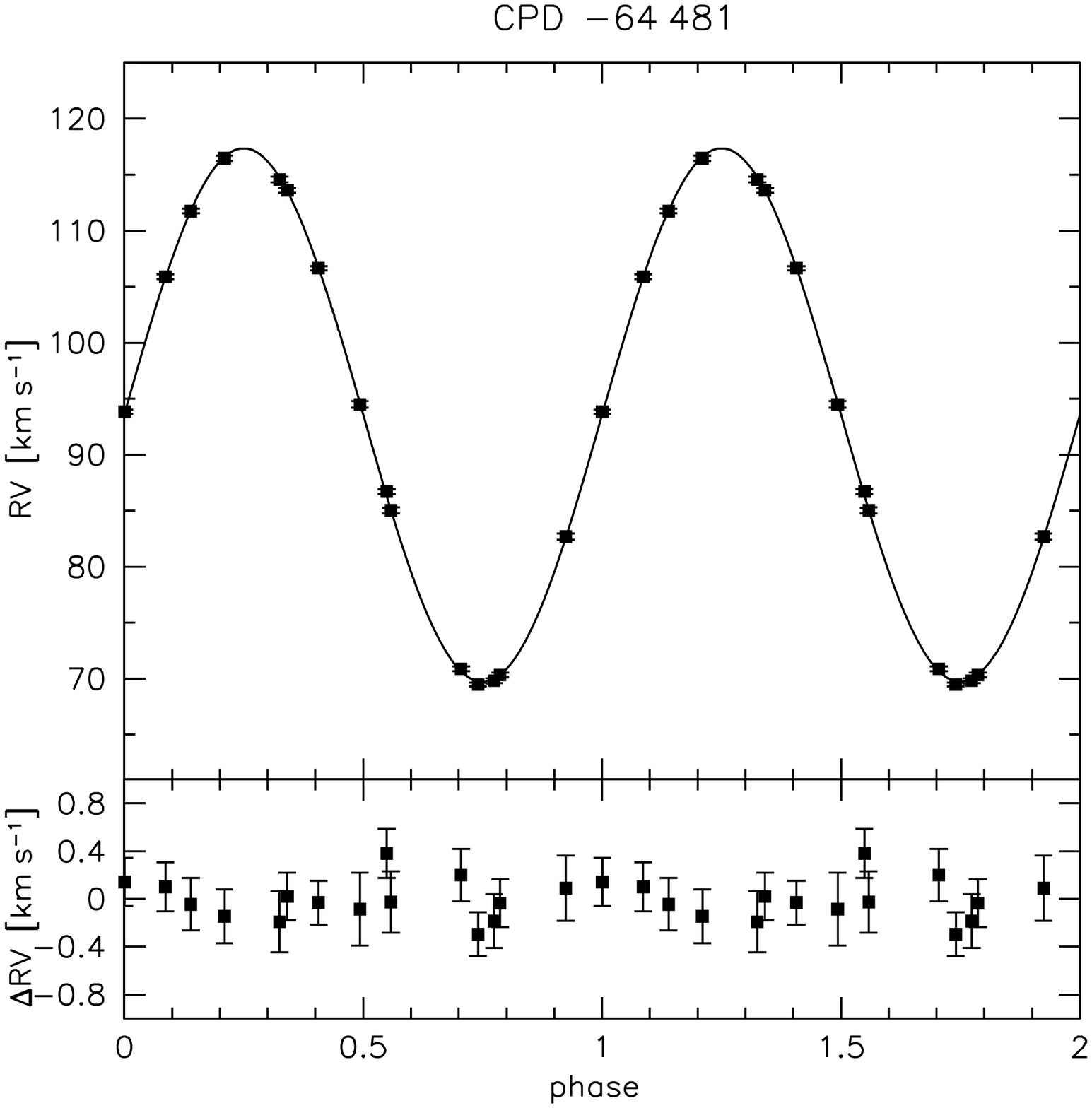}}

\caption{Radial velocity plotted against orbital phase. The RV data were phase folded with the most likely orbital periods. The residuals are plotted below.}
\label{rv1}
\end{figure*}

The discovery of a brown dwarf ($M_{\rm BD}=0.053\pm0.006\,M_{\rm \odot}$) in close orbit ($0.08\,{\rm d}$) around such a white dwarf supports this scenario and shows that substellar companions can influence late stellar evolution (Maxted et al. \cite{maxted06}). With the discovery of the eclipsing sdB+BD binaries SDSS\,J0820+0008 and SDSSJ\,J1622+4730 in the course of the MUCHFUSS project (Geier et al. \cite{geier11a}), it was shown observationally that substellar companions are also able to form sdBs (Geier et al. \cite{geier11c}; Schaffenroth et al. \cite{schaffenroth14}). 

We now have to address the question, how massive the companion must be to survive the CE-phase. It might also be possible to form an sdB and either evaporate the substellar companion or merge it with the red-giant core. Substellar companions of brown dwarf and even planetary mass in wide orbits have been detected around pulsating sdBs (Silvotti et al. \cite{silvotti07}; Lutz et al. \cite{lutz12}) and eclipsing sdB binaries using the timing method (see Zorotovic \& Schreiber \cite{zorotovic13} and references therein for a summary). Those results, although still under debate (see Wittenmyer et al. \cite{wittenmyer}, Horner et al. \cite{horner}), suggest that a high fraction of the sdB stars might be orbited by such objects. Substellar companions in close orbit might therefore be frequent as well.

Here we report the discovery of a reflection effect, but no eclipses, in the light curves of two close sdB binaries. CPD-64$^{\circ}$481 and PHL\,457 have been reported to be close sdB binaries with small RV shifts by Edelmann et al. (\cite{edelmann05}). Furthermore, PHL\,457 has been identified as long-period pulsator of V\,1093\,Her type (Blanchette et al. \cite{blanchette08}). Those two sdBs are among the best studied close sdB binaries. Detailed analyses showed that both are normal sdB binaries with typical atmospheric parameters (CPD-64$^{\circ}$481: $T_{\rm eff}= 27 500\pm 500\,\rm K,\, \log{g}=5.60 \pm 0.05$, Geier et al. \cite{geier10}; PHL\,457: $T_{\rm eff}= 26 500\pm 500\,\rm K,\, \log{g}=5.38\pm 0.05$, Geier et al. \cite{geier13a}).

Geier at al. (\cite{geier10}) constrained the companion mass of CPD-64$^{\circ}$481 to be as high as 0.62 $M_{\rm \odot}$ by measuring the projected rotational velocity of the sdB and assuming synchronised rotation. This assumption is reasonable, as the theoretical synchronisation timescales with stellar mass companions due to tidal interactions for binaries with periods of about 0.3 d are much smaller or comparable to the lifetime of the sdB on the EHB, depending on the theory (see Geier et al. \cite{geier10}).
The inclination angle was predicted to be as small as $7^{\rm \circ}$. Because no traces of the companion are seen in the spectrum, they concluded that the companion must be a WD, as main sequence stars would be visible in the optical spectra, if their masses are higher than $\sim0.45\,M_{\rm \odot}$ (Lisker et al. \cite{lisker05}). However, the detection of the reflection effect rules out such a compact companion. 

Using the same method we constrained the companion mass of PHL\,457. Although the companion mass assuming synchronisation ($\sim0.26\,M_{\rm \odot}$) would still be consistent with observations, the derived inclination angle of $8^{\rm \circ}$ is very small and therefore unlikely. 

Moreover, observational evidence, both from asteroseismic studies (Pablo et al. \cite{pablo11,pablo12}) and spectroscopic measurements (Schaffenroth et al. \cite{schaffenroth14}), indicates that synchronisation is not generally established in sdB binaries with low-mass companions (see also the discussion in Geier et al. \cite{geier10}). Therefore, the rotation of the sdBs in CPD-64$^{\circ}$481 and PHL\,457 is most likely not synchronised with their orbital motion and the method described in Geier et al. (\cite{geier10}) not applicable.


\section{Time-resolved spectroscopy and orbital parameters}\label{s:data}

In total, 45 spectra were taken with the FEROS spectrograph ($R\simeq48000,\lambda=3800-9200\,{\rm \AA}$) mounted at the ESO/MPG-2.2m telescope for studies of sdB stars at high resolution (Edelmann et al. \cite{edelmann05}; Geier et al. \cite{geier10}; Classen et al. \cite{classen11}). The spectra were reduced with the FEROS pipeline available in the \texttt{MIDAS} package. The FEROS pipeline, moreover, performs the barycentric correction.   

To measure the RVs with high accuracy, we chose a set of sharp, unblended metal lines situated between $3600\,{\rm \AA}$ and $6600\,{\rm \AA}$. Accurate rest wavelengths were taken from the NIST database. Gaussian and Lorentzian profiles were fitted using the SPAS (Hirsch \cite{hirsch}) and FITSB2 routines (Napiwotzki et al. \cite{napiwotzki04b}, for details see Classen et al. \cite{classen11}). To check the wavelength calibration for systematic errors we used telluric features as well as night-sky emission lines. The FEROS instrument turned out to be very stable. Usually corrections of less than $0.5\,{\rm km\,s^{-1}}$ had to be applied. The RVs and formal $1\sigma$-errors are given in Appendix~\ref{app:RV}.

The orbital parameters and associated false-alarm probabilities are determined as described in Geier et al. (\cite{geier11c}). In order to estimate the significance of the orbital solutions and the contributions of systematic effects to the error budget, we normalised the $\chi^{2}$ of the most probable solution by adding systematic errors $e_{\rm norm}$ in quadrature until the reduced $\chi^{2}$ reached $\simeq1.0$. The hypothesis that both orbital periods are correct can be accepted with a high degree of confidence. The phased RV curves for the best solutions are of excellent quality (see Fig.~\ref{rv1}). The derived orbital parameters are given in Table~\ref{tab:orbits} and the orbital solution for CPD-64$^{\circ}$481 is perfectly consistent with the result presented in Edelmann et al. (\cite{edelmann05}).

\begin{table*}[t!]
\caption{Derived orbital solutions, mass functions and minimum companion masses} 
\label{tab:orbits}
\centering
\begin{tabular}{lllrrlllll} 
Object & $T_{0}^a$ & $\rm P^a$ &  $\gamma^a$ & $\rm K^a$ & $e_{\rm norm}$ & $f(M)$ & $M_{\rm 2}^b$ & $i_{\rm max}^c$\\
 & [$-$2\,450\,000] & [d] & [${\rm km\,s^{-1}}$] & [${\rm km\,s^{-1}}$] & [${\rm km\,s^{-1}}$] & [$M_{\rm \odot}$] & [$M_{\rm \odot}$] &$^\circ$\\ 
\hline
\\[-3mm]
CPD-64$^{\circ}$481 & $3431.5796\pm0.0002$ & $0.27726315\pm0.00000008$ & $93.54\pm0.06$ & $23.81\pm0.08$ & $0.16$ & $0.0004$ & $>0.048$ & 70\\
PHL\,457 & $5501.5961\pm0.0009$ & $0.3131\pm0.0002$   & $20.7\pm0.2$ & $13.0\pm0.2$ & $0.7$ & $0.00007$ & $>0.027$ & 75 \\
\hline \\[-3mm]
\end{tabular}
\begin{list}{}{} 
\item[$^{\mathrm{a}}$] The systematic error adopted to normalise the reduced $\chi^{2}$ ($e_{\rm norm}$) is given for each case.
\item[$^{\mathrm{b}}$] The minimum companion masses take into account the highest possible inclination.
\item[$^{\mathrm{c}}$] $i=90$ is defined as an edge-on orbit
\end{list}
 
\end{table*}

\section{Photometry}

Time-resolved differential photometry in BVR-filters for CPD-64$^{\circ}$481 and VR filters for PHL\,457 was obtained with the SAAO STE4 CCD
on the 1.0m telescope at the Sutherland site of the South African Astronomical Observatory (SAAO). Photometric reductions 
were performed using an automated version of DOPHOT (Schechter et al. \cite{schechter93}).

\begin{figure*}[t!]
\centering
	\resizebox{9.15cm}{!}{\includegraphics[angle=-90]{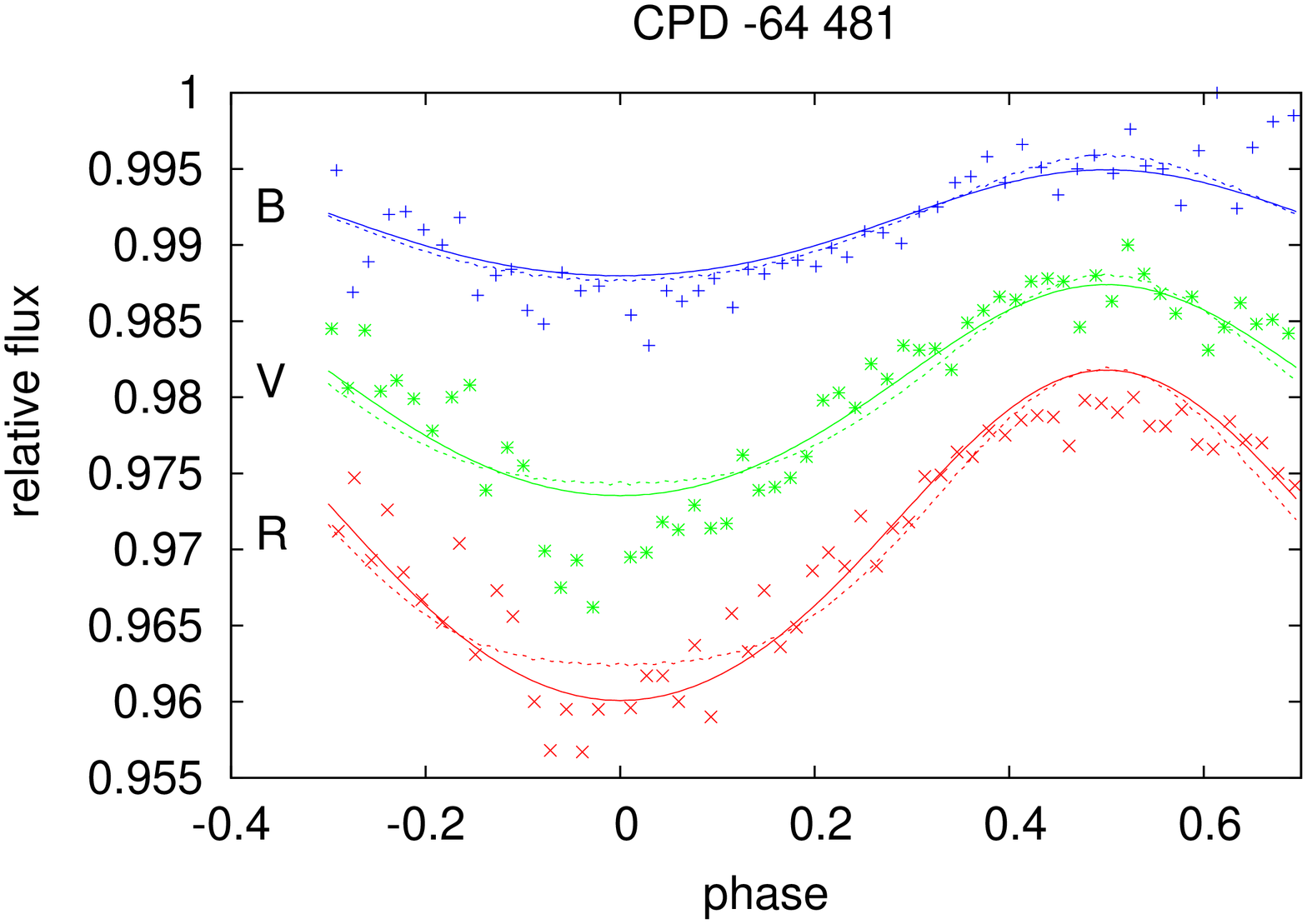}}
        \resizebox{9.15cm}{!}{\includegraphics[angle=-90]{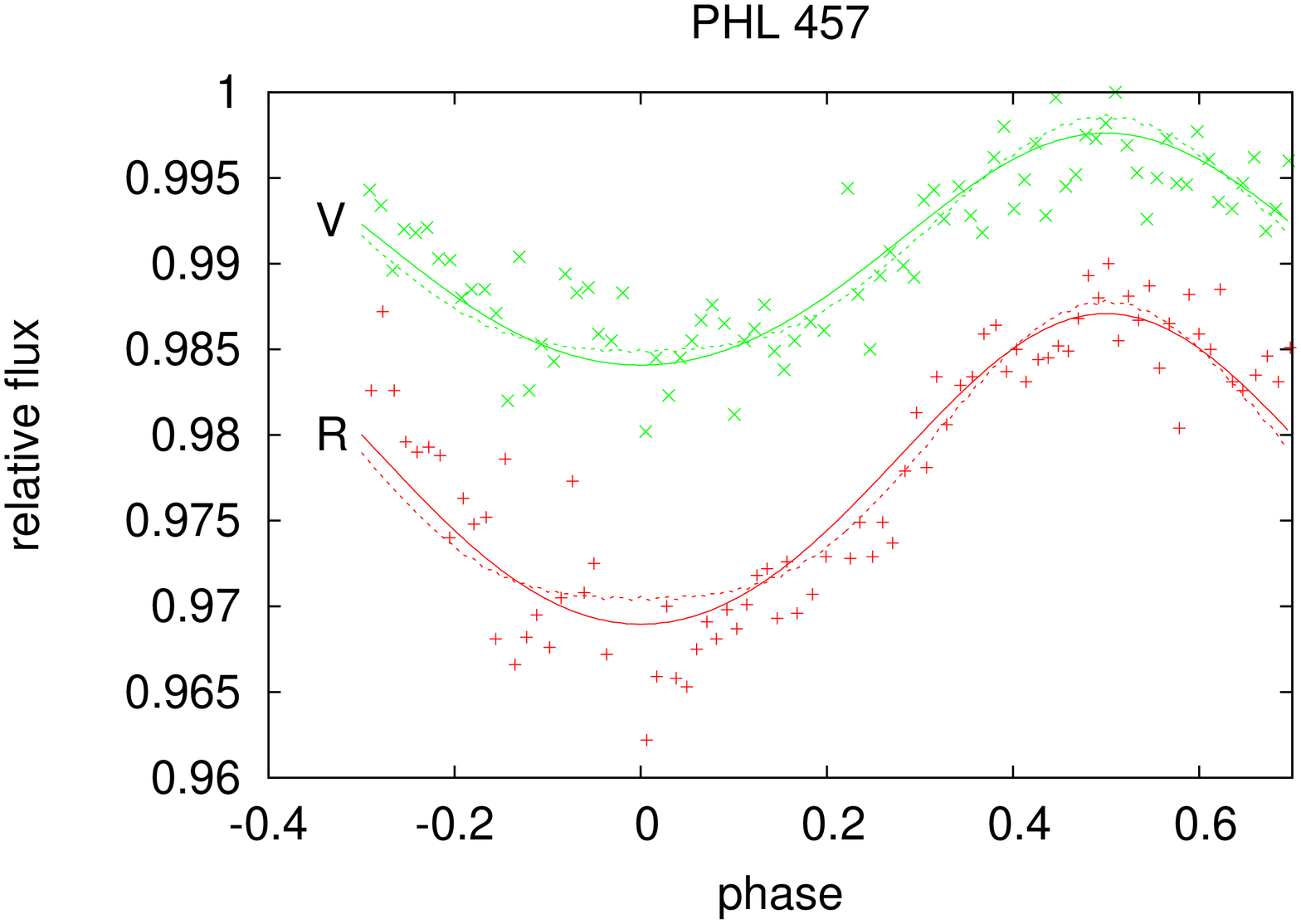}}
	\caption{Phased and binned light curves in B-, V- and R-bands in the case of CPD-64$^{\circ}$481 (left panel) and V- and R-bands in the case of PHL\,457 (right panel).  Overplotted are two models for an inclination of 10$^\circ$ (solid) and 65$^\circ$ (dashed) for CPD-64$^{\circ}$481 and 10$^\circ$ (solid) and 70$^\circ$(dashed) for PHL\,457. The lightcurve models with higher inclinations (dashed) have broader minima and shallower maxima.}
	\label{lc}

\end{figure*}

\begin{figure*}[h!]
\centering
	\resizebox{9.15cm}{!}{\includegraphics{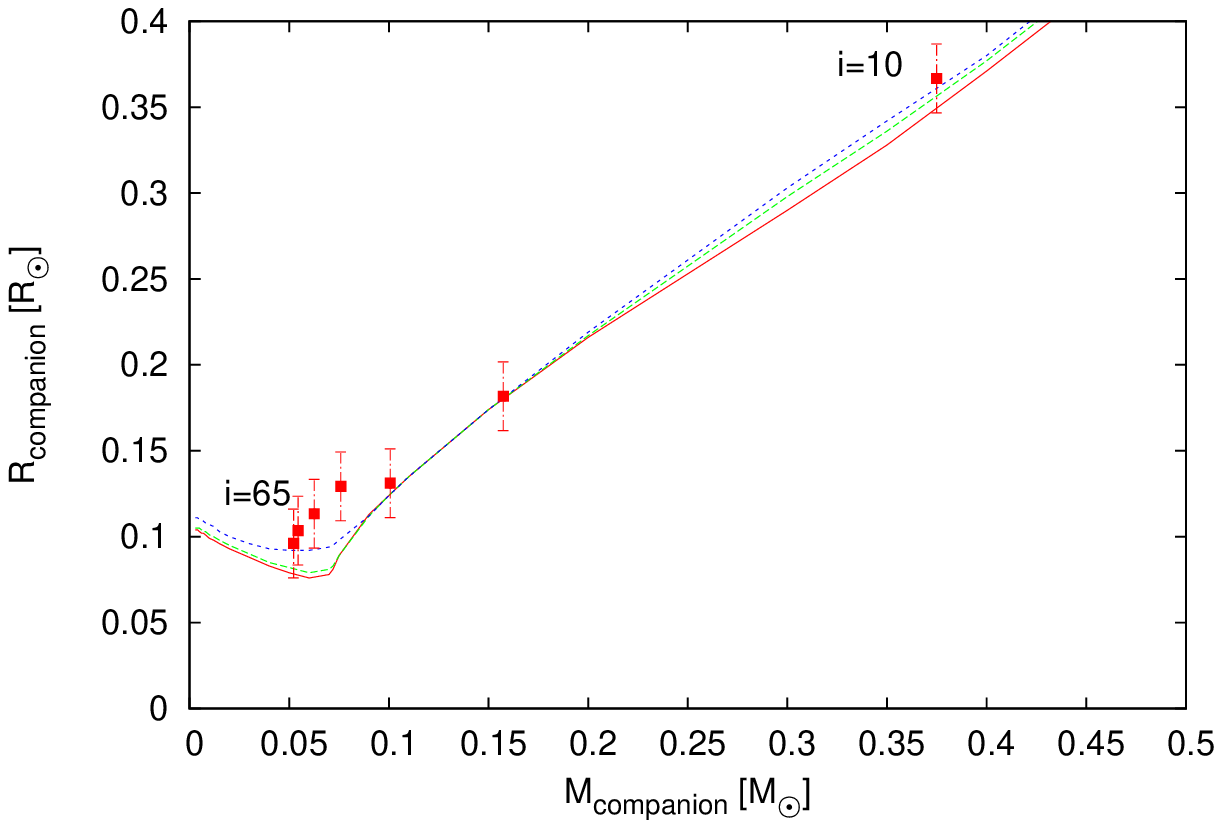}}
	\resizebox{9.15cm}{!}{\includegraphics{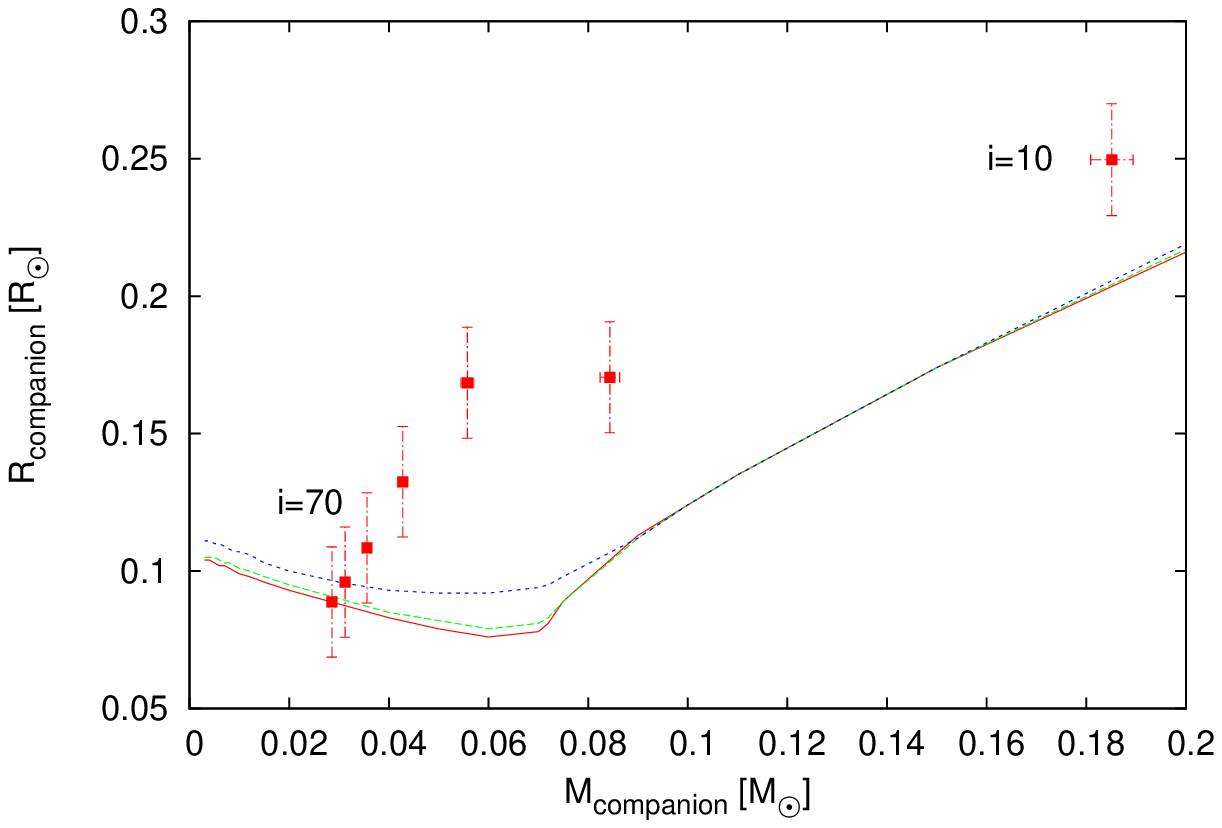}}
		\caption{Mass-radius relation of the companion of CPD-64$^{\circ}$481 (left panel) and PHL\,457 (right panel) for different inclinations (filled rectangles), compared to theoretical relations by Chabrier \& Baraffe (\cite{chabrier97}) for an age of the system of 1Gyr (short dashed line) , 5 Gyr (long dashed line) and 10 Gyr (solid line). For all shown solutions mass and radius of the sdB are also consistent with the spectroscopic surface gravity measurement of $\log{g}=5.60 \pm 0.05$ for CPD-64$^{\circ}$481 (Geier et al. \cite{geier10}) and $\log{g}=5.38\pm 0.05$ (Geier et al. \cite{geier13a}) for PHL\,457.}
		\label{m-r_cpd}

\end{figure*}

The differential light curves have been phased to the orbital periods derived from the RV-curves and 
binned to achieve higher S/N. The light curves show sinusoidal variations ($\sim10\,{\rm mmag}$) with orbital phase characteristic for 
a reflection effect (Fig.~\ref{lc}). It originates from the irradiation of a cool companion by the hot subdwarf primary. 
The projected area of the companion's heated hemisphere changes while it orbits the primary. Compared to other reflection effect binaries the amplitude of the reflection effect in both systems is quite small. The amplitude of the reflection effect depends mostly on the separation of the system, the effective temperature of the subdwarf, and the visible irradiated area of the companion. Seen edge-on, the relative change of this area is the highest. However, for small inclinations the derived mass of the companion becomes higher and because there is a strong correlation between mass and radius on the lower main sequence (see Fig. \ref{m-r_cpd}), the radius of the companion and the absolute irradiated area becomes larger as well. Due to this degeneracy it is therefore not straight forward to claim that small reflection effects can simply be explained by small inclination angles.


We fitted models calculated with MORO, which is based on the Wilson-Devinney code (MOdified ROche model, Drechsel et al. \cite{drechsel95}), to the light curves as described in Schaffenroth et al. (\cite{schaffenroth13}). As no eclipses are present, the inclination is difficult to determine and we fitted light curve solutions for different fixed inclinations. The mass ratio, which can be calculated from the mass function for different inclinations, was also kept fixed, so that the mass of the sdB is equal to the canonical sdB mass $M_{\rm sdB}=0.47\,{\rm M_{\odot}}$ (see Fontaine et al. \cite{fontaine12} and references therein). Shape and amplitude of the variation mostly depends on the orbital inclination and the mass ratio, but also on the radius ratio of both components and the unknown albedo of the companion. Due to this high number of parameters, that are not independent from each other, we cannot find a unique solution.
 
Selecting only solutions for which the photometric radius is consistent with the spectroscopic radius derived from the surface gravity, we narrow down the number of solutions. Unfortunately, due to the degeneracy between the binary inclination and the radius of the companion we find equally good solutions for each inclination (see also \O stensen et al. \cite{oestensen13}). In the case of CPD-64$^{\circ}$481, see Fig.~\ref{m-r_cpd}, the derived mass and radius of the companion are in agreement with theoretical relations by Chabrier \& Baraffe (\cite{chabrier97}) for the whole range of inclinations. In the case of PHL\,457 the theoretical mass-radius relation is only consistent for an inclination of 50-70$^\circ$. For lower inclinations the measured radius would be larger than expected by the models. However, due to the many assumptions used in the analysis, it is difficult to estimate the significance of this result, as a smaller mass for the sdB could solve this issue. 
  
In Fig. \ref{lc} we show model light curves for high and low inclination. Although small differences are present, a much better quality light curve is needed to resolve them. The sum of the deviations between the measurements and the models are somewhat, but not significantly, smaller for low inclinations. Therefore, we cannot draw firm conclusions from our photometric data at hand. 

\section{Brown dwarf nature of the unseen companions}

The two binaries are single-lined and their binary mass functions $f_{\rm m} = M_{\rm comp}^3 \sin^3i/(M_{\rm comp} + M_{\rm sdB})^2 = P K^3/2 \pi G$ can be determined.
The RV semi-amplitude and the orbital period can be derived from the RV curve, but the sdB mass $M_{\rm sdB}$, the companion mass $M_{\rm comp}$ and the inclination angle $i$ remain free parameters. Adopting the canonical sdB mass $M_{\rm sdB}=0.47\,{\rm M_{\odot}}$ and the $i_{\rm max}$, that can be constrained, because no eclipses are present in the lightcurve, we derive lower limits for the companion masses (see Table\,\ref{tab:orbits}). 

Those minimum masses of $0.048\,M_{\rm \odot}$ for CPD-64$^{\circ}$481 and $0.027\,M_{\rm \odot}$ for PHL\,457 - the smallest minimum companion mass measured in any sdB binary so far - are significantly below the hydrogen-burning limit ($\sim0.07-0.08\,M_{\rm \odot}$, Chabrier et al. \cite{chabrier00}). As no features of the companion are found in the spectrum, we also derive an upper mass limit of $\sim0.45\,M_{\rm \odot}$ (Lisker et al. \cite{lisker05}) .



The initial sample of Edelmann et al. (\cite{edelmann05}) consisted of known, bright hot subdwarf stars. Because no additional selection criteria were applied, it can be assumed that the inclination angles of the binaries found in this survey are randomly distributed. Due to the projection effect it is much more likely to find binary systems at high rather than low  inclinations. The probability, that a binary has an inclination higher than a certain angle, can be calculated as described in Gray (\cite{gray92}), $P_{i>i_0}=1-(1-\cos i_0)$ . Since the companion mass scales with the inclination angle, we can derive the probability that the mass of the companion is smaller than the hydrogen-burning limit of $\sim0.08\,M_{\rm \odot}$, which separates stars from brown dwarfs.

For CPD-64$^{\circ}$481, the inclination angle must be higher than $38^{\rm \circ}$, which translates into a probability of $79\%$. In the case of PHL\,457, an inclination higher than $21^{\rm \circ}$ is required and the probability for the companion to be a brown dwarf is as high as $94\%$. 
We therefore conclude that the cool companions in those two binary systems are likely brown dwarfs.

The only chance to constrain the inclination better would be very high signal-to-noise lightcurves.
Moreover, high resolution, high S/N spectra could help to constrain the mass ratio of the system. They could allow to discover emission lines from the irradiated hemisphere of the companion, as done for the sdOB+dM system AA Dor (Vu{\v c}kovi\'c et al. 2008). The strength of the emission lines should be independent of the inclination, depending only on the size of the companion, the separation of the system and the effective temperature of the primary. As the systems are very bright, it might be possible to find these emission lines despite the larger separation and lower effective temperature of our systems.

\begin{figure}[t!]
\centering
	\resizebox{9.15cm}{!}{\includegraphics{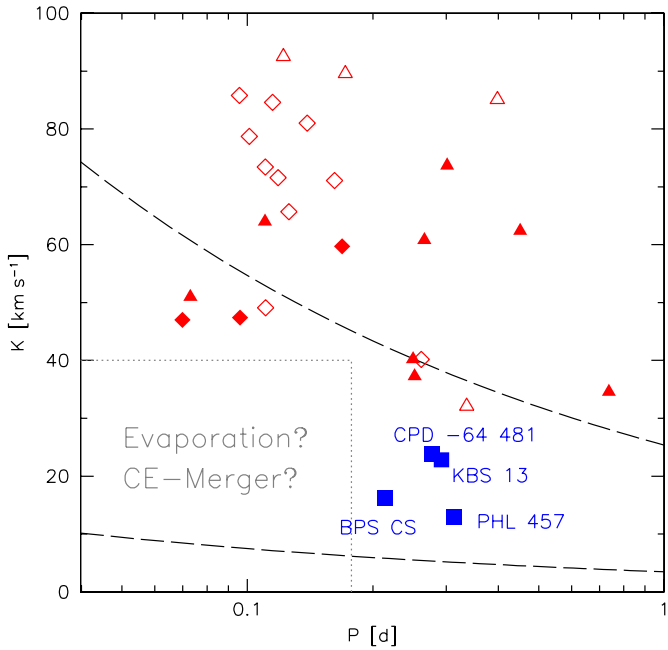}}
        \caption{The RV semiamplitudes of all known sdB binaries with reflection effects and spectroscopic solutions plotted against their orbital periods (Kupfer et al. \cite{kupfer13}). Diamonds mark eclipsing sdB binaries of HW\,Vir type where the companion mass is well constrained, triangles systems without eclipses, where only lower limit can be derived for the companion masses. Squares mark CPD-64$^{\circ}$481, PHL\,457, KBS\,13 and BPS\,CS\,22169$-$0001. Open symbols mark systems that have been discovered based on photometry, filled symbols have been discovered based on spectroscopy. The dashed lines mark the regions to the right where the minimum companion masses derived from the binary mass function (assuming $0.47\,M_{\rm \odot}$ for the sdBs) exceed $0.01\,M_{\rm \odot}$ (lower curve) and $0.08\,M_{\rm \odot}$ (upper curve).}
        \label{periodK}

\end{figure}
\section{Discussion}

Figure~\ref{periodK} gives an overview of the 29 sdB binaries with reflection effect and known orbital parameters (Kupfer et al. \cite{kupfer13}). While most companions are late M-dwarfs with masses close to $\sim0.1\,M_{\rm \odot}$, there is no sharp drop below the hydrogen-burning limit. The fraction of close substellar companions is substantial. An obvious feature in Fig.~\ref{periodK} is the lack of binaries with periods shorter than $\sim0.2\,{\rm d}$ and $K<50\,{\rm km\,s^{-1}}$ corresponding to companion masses of less than $\sim0.06\,M_{\rm \odot}$. 

This feature could not be due to selection effects. About half of the known reflection effect binaries have been found based on RV-shifts detected in time-resolved spectra. As has been shown in this work, RV-semiamplitudes of a few tens of ${\rm km\,s^{-1}}$ are easily measurable. Furthermore, short-period binaries are found and solved easier than long-period systems.

The other half of the sample has been discovered based on variations in their light curves. Shape and amplitude of the light curves depend mostly on the radius of the companion for similar orbital periods and separations. Since the radii of late M-dwarfs, brown dwarfs and also Jupiter-size planets are very similar ($\sim0.1\,R_{\rm \odot}$), their light curves are expected to be very similar as well. 

The most likely reason for this gap is the merger or evaporation of low-mass companions either before or after the CE-ejection corresponding to a population of single sdB stars. Other recent discoveries are perfectly consistent with this scenario. Charpinet et al. (\cite{charpinet11}) reported the discovery of two Earth-sized bodies orbiting a single pulsating sdB within a few hours. These might be the remnants of a more massive companion evaporated in the CE-phase (Bear \& Soker \cite{bear12}). Geier et al. (\cite{geier11b,geier13b}) found two fast rotating single sdBs, which might have formed in a CE-merger. Those discoveries provide further evidence that substellar companions play an important role in the formation of close binary and likely also single sdB stars. 

We therefore conclude that the lack of short period systems with small RV variations $\rm K<50\,{\rm km\,s^{-1}}$ is real. However, the probability that substellar companions are present in systems with longer periods ($>0.2\,{\rm d}$) is quite high. In addition to the two binaries discussed here, two more systems with similar orbital parameters and reflection effects have been found (KBS\,13, For et al. \cite{for}; BPS\,CS\,22169$-$0001, Geier et al. \cite{geier12}). Following the line of arguments outlined above, we calculate the probability for those two systems to host a stellar companion to be $9\%$ for BPS\,CS\,22169$-$0001 and $20\%$ for KBS\,13. Multiplying those numbers  for all four binaries, we conclude that the probability that none of them has a substellar companions is less than $0.02\%$. 

\begin{acknowledgements}
Based on observations at the La Silla Observatory of the European Southern Observatory for programmes number 073.D-0495(A), 074.B-455(A) and 086.D-0714(A). This paper uses observations made at the South African Astronomical Observatory (SAAO). V.S. acknowledges funding by the Deutsches Zentrum f\"ur Luft- und Raumfahrt (grant 50 OR 1110) and by the Erika-Giehrl-Stiftung.

\end{acknowledgements}

\newpage

\begin{appendix}

\vspace{10cm}
\section{Radial velocities}\label{app:RV}

\begin{table}[h!]
\caption{CPD-64$^{\circ}$481: All spectra were acquired with
the FEROS instrument}
\label{RV1}
\centering
\begin{tabular}{lr}
\hline
\noalign{\smallskip}
mid$-$HJD & RV [${\rm km\,s^{-1}}$]\\
$-2\,450\,000$ & \\
\noalign{\smallskip}
\hline
\noalign{\smallskip}
3249.89041 &  70.87   $\pm$ 0.15 \\
3250.89408 &  114.58  $\pm$ 0.20 \\
3251.85033 &  69.81   $\pm$ 0.16 \\
3252.88158 &  94.51   $\pm$ 0.26 \\
3252.89956 &  85.03   $\pm$ 0.20 \\
3253.83281 &  82.69   $\pm$ 0.22 \\
3253.87777  & 105.90  $\pm$ 0.13 \\
3253.91205  & 116.47  $\pm$ 0.16 \\
3425.51834  & 111.76  $\pm$ 0.15 \\
3426.51685  & 69.48   $\pm$ 0.09 \\
3427.53343  & 106.65  $\pm$ 0.09 \\
3428.52983  & 93.84   $\pm$ 0.12 \\
3429.51370  & 86.71   $\pm$ 0.13 \\
3430.56510  & 113.59  $\pm$ 0.12 \\
3431.52045  & 70.33   $\pm$ 0.12 \\
\noalign{\smallskip}
\hline
\end{tabular}

\end{table}

\begin{table}[h!]
\caption{PHL\,457: All spectra were acquired with
the FEROS instrument }
\label{RV5}
\centering
\begin{tabular}{lrl}
\hline
\noalign{\smallskip}
mid$-$HJD & RV [${\rm km\,s^{-1}}$]\\
$-2\,450\,000$ &\\
\noalign{\smallskip}
\hline
\noalign{\smallskip}
3249.64149     & 10.1    $\pm$ 0.2\\
3250.64322     & 22.2    $\pm$ 0.2\\
3251.58746     & 23.8    $\pm$ 0.3\\
3253.56948     & 29.4    $\pm$ 0.3\\
5500.52235     & 15.1    $\pm$ 0.2\\
5500.54504     & 9.7     $\pm$ 0.5\\
5501.52731     & 8.4     $\pm$ 0.2\\
5501.53727     & 9.1     $\pm$ 0.3\\
5501.54722     & 9.3     $\pm$ 0.2\\
5501.55718     & 11.0    $\pm$ 0.1\\
5501.56712     & 14.1    $\pm$ 0.2\\
5501.57705     & 16.4    $\pm$ 0.2\\
5501.58698     & 17.2    $\pm$ 0.2\\
5501.59692     & 21.1    $\pm$ 0.3\\
5501.60685     & 22.2    $\pm$ 0.8\\
5502.50148     & 13.0    $\pm$ 0.4\\
5502.51028     & 13.9    $\pm$ 0.2\\
5502.51908     & 16.1    $\pm$ 0.3\\
5502.52785     & 18.5    $\pm$ 0.3\\
5502.53663     & 21.8    $\pm$ 0.3\\
5502.54542     & 23.4    $\pm$ 0.3\\
5502.55420     & 24.3    $\pm$ 0.3\\
5502.56298     & 28.4    $\pm$ 0.3\\
5502.57175     & 30.9    $\pm$ 0.3\\
5502.58053     & 31.0    $\pm$ 0.4\\
5502.58932     & 31.3    $\pm$ 0.4\\
5502.59811     & 33.2    $\pm$ 0.4\\
5502.60688     & 34.8    $\pm$ 0.4\\
5502.61566     & 34.1    $\pm$ 0.4\\
5502.62502     & 32.7    $\pm$ 0.3\\
\noalign{\smallskip}
\hline
\end{tabular}

\end{table}
\newpage


\end{appendix}

\end{document}